\documentclass{emulateapj}

\slugcomment{Draft Version}

\shorttitle{MULTI-TeV GAMMA-RAY OBSERVATION FROM THE CRAB}
\shortauthors{Amenomori et al.}

\begin{document}


\title{MULTI-TeV GAMMA-RAY OBSERVATION FROM THE CRAB NEBULA \\
USING THE TIBET-III AIR SHOWER ARRAY \\
FINELY TUNED BY THE COSMIC-RAY MOON'S SHADOW}


\author{
M.~Amenomori\altaffilmark{1}, X.~J.~Bi\altaffilmark{2}, D.~Chen\altaffilmark{3}, S.~W.~Cui\altaffilmark{4},
Danzengluobu\altaffilmark{5}, L.~K.~Ding\altaffilmark{2}, X.~H.~Ding\altaffilmark{5}, C.~Fan\altaffilmark{6},
C.~F.~Feng\altaffilmark{6}, Zhaoyang Feng\altaffilmark{2}, Z.~Y.~Feng\altaffilmark{7},
X.~Y.~Gao\altaffilmark{8}, Q.~X.~Geng\altaffilmark{8}, H.~W.~Guo\altaffilmark{5}, H.~H.~He\altaffilmark{2},
M.~He\altaffilmark{6}, K.~Hibino\altaffilmark{9}, N.~Hotta\altaffilmark{10}, Haibing~Hu\altaffilmark{5},
H.~B.~Hu\altaffilmark{2}, J.~Huang\altaffilmark{2,3}, Q.~Huang\altaffilmark{7}, H.~Y.~Jia\altaffilmark{7},
F.~Kajino\altaffilmark{11}, K.~Kasahara\altaffilmark{12}, Y.~Katayose\altaffilmark{13},
C.~Kato\altaffilmark{14}, K.~Kawata\altaffilmark{3}, Labaciren\altaffilmark{5}, G.~M.~Le\altaffilmark{15},
A.~F.~Li\altaffilmark{6}, J.~Y.~Li\altaffilmark{6}, Y.-Q.~Lou\altaffilmark{16}, H.~Lu\altaffilmark{2},
S.~L.~Lu\altaffilmark{2}, X.~R.~Meng\altaffilmark{5}, K.~Mizutani\altaffilmark{12,17}, J.~Mu\altaffilmark{8},
K.~Munakata\altaffilmark{14}, A.~Nagai\altaffilmark{18}, H.~Nanjo\altaffilmark{1},
M.~Nishizawa\altaffilmark{19}, M.~Ohnishi\altaffilmark{3}, I.~Ohta\altaffilmark{20},
H. Onuma\altaffilmark{17}, T.~Ouchi\altaffilmark{9}, S.~Ozawa\altaffilmark{12}, J.~R.~Ren\altaffilmark{2},
T.~Saito\altaffilmark{21}, T.~Y.~Saito\altaffilmark{22}, M.~Sakata\altaffilmark{11},
T.~K.~Sako\altaffilmark{3}, M.~Shibata\altaffilmark{13}, A.~Shiomi\altaffilmark{23},
T.~Shirai\altaffilmark{9}, H.~Sugimoto\altaffilmark{24}, M.~Takita\altaffilmark{3},
Y.~H.~Tan\altaffilmark{2}, N.~Tateyama\altaffilmark{9}, S.~Torii\altaffilmark{12},
H.~Tsuchiya\altaffilmark{25}, S.~Udo\altaffilmark{12}, B.~Wang\altaffilmark{2}, H.~Wang\altaffilmark{2},
X.~Wang\altaffilmark{12}, Y.~Wang\altaffilmark{2}, Y.~G.~Wang\altaffilmark{6}, H.~R.~Wu\altaffilmark{2},
L.~Xue\altaffilmark{6}, Y.~Yamamoto\altaffilmark{11}, C.~T.~Yan\altaffilmark{3}, X.~C.~Yang\altaffilmark{8},
S.~Yasue\altaffilmark{26}, Z.~H.~Ye\altaffilmark{15}, G.~C.~Yu\altaffilmark{7}, A.~F.~Yuan\altaffilmark{5},
T.~Yuda\altaffilmark{9}, H.~M.~Zhang\altaffilmark{2}, J.~L.~Zhang\altaffilmark{2},
N.~J.~Zhang\altaffilmark{6}, X.~Y.~Zhang\altaffilmark{6}, Y.~Zhang\altaffilmark{2}, Yi~Zhang\altaffilmark{2},
Zhaxisangzhu\altaffilmark{5} and X.~X.~Zhou\altaffilmark{7} \\
(The Tibet AS$\gamma$ Collaboration)
}

\altaffiltext{1}{Department of Physics, Hirosaki University, Hirosaki 036-8561, Japan.}
\altaffiltext{2}{Key Laboratory of Particle Astrophysics, Institute of High Energy Physics, Chinese Academy of Sciences, Beijing 100049, China.}
\altaffiltext{3}{Institute for Cosmic Ray Research, University of Tokyo, Kashiwa 277-8582, Japan.}
\altaffiltext{4}{Department of Physics, Hebei Normal University, Shijiazhuang 050016, China.}
\altaffiltext{5}{Department of Mathematics and Physics, Tibet University, Lhasa 850000, China.}
\altaffiltext{6}{Department of Physics, Shandong University, Jinan 250100, China.}
\altaffiltext{7}{Institute of Modern Physics, SouthWest Jiaotong University, Chengdu 610031, China.}
\altaffiltext{8}{Department of Physics, Yunnan University, Kunming 650091, China.}
\altaffiltext{9}{Faculty of Engineering, Kanagawa University, Yokohama 221-8686, Japan.}
\altaffiltext{10}{Faculty of Education, Utsunomiya University, Utsunomiya 321-8505, Japan.}
\altaffiltext{11}{Department of Physics, Konan University, Kobe 658-8501, Japan.}
\altaffiltext{12}{Research Institute for Science and Engineering, Waseda University, Tokyo 169-8555, Japan.}
\altaffiltext{13}{Faculty of Engineering, Yokohama National University, Yokohama 240-8501, Japan.}
\altaffiltext{14}{Department of Physics, Shinshu University, Matsumoto 390-8621, Japan.}
\altaffiltext{15}{Center of Space Science and Application Research, Chinese Academy of Sciences, Beijing 100080, China.}
\altaffiltext{16}{Physics Department and Tsinghua Center for Astrophysics, Tsinghua University, Beijing 100084, China.}
\altaffiltext{17}{Department of Physics, Saitama University, Saitama 338-8570, Japan.}
\altaffiltext{18}{Advanced Media Network Center, Utsunomiya University, Utsunomiya 321-8585, Japan.}
\altaffiltext{19}{National Institute of Informatics, Tokyo 101-8430, Japan.}
\altaffiltext{20}{Sakushin Gakuin University, Utsunomiya 321-3295, Japan.}
\altaffiltext{21}{Tokyo Metropolitan College of Industrial Technology, Tokyo 116-8523, Japan.}
\altaffiltext{22}{Max-Planck-Institut f\"ur Physik, M\"unchen D-80805, Deutschland.}
\altaffiltext{23}{College of Industrial Technology, Nihon University, Narashino 275-8576, Japan.}
\altaffiltext{24}{Shonan Institute of Technology, Fujisawa 251-8511, Japan.}
\altaffiltext{25}{RIKEN, Wako 351-0198, Japan.}
\altaffiltext{26}{School of General Education, Shinshu University, Matsumoto 390-8621, Japan.}




\begin{abstract}
The Tibet-III air shower array, consisting of 533 scintillation
detectors, has been operating successfully at Yangbajing in Tibet, China since
1999.  Using the dataset collected by this array from 1999 November
through 2005 November, we obtained the energy spectrum of
$\gamma$-rays from the Crab Nebula, expressed by a power law
as $(dJ/dE) = (2.09\pm0.32)\times10^{-12} (E/{\rm
  3~TeV})^{-2.96\pm0.14} {\rm cm}^{-2} {\rm s}^{-1} {\rm TeV}^{-1}$ in
the energy range of 1.7 to 40~TeV. This result is consistent
with other independent $\gamma$-ray observations by imaging air Cherenkov
telescopes. In this paper, we carefully checked and tuned the
performance of the Tibet-III array using data on the moon's shadow in
comparison with a detailed Monte Carlo simulation. 
The shadow is shifted to the west of the moon's apparent position as an effect
of the geomagnetic field, although the extent of this displacement depends 
on the primary energy positively charged cosmic rays. This finding enables us to estimate
the systematic error in determining the primary energy from its
shower size. This error is estimated to be less than $\pm$12\% in
our experiment.  This energy scale estimation is the first attempt among cosmic-ray
experiments at ground level.  The systematic pointing error is also
estimated to be smaller than $0\fdg011$. The deficit rate and position
of the moon's shadow are shown to be very stable within a statistical
error of $\pm$6\% year by year.  This guarantees the long-term stability
of point-like source observation with the Tibet-III array. These systematic errors are
adequately taken into account in our study of the Crab Nebula.
\end{abstract}





\section{INTRODUCTION}

The Crab~Nebula is a standard source of
radiation in the northern sky across a wide energy band, from radio to near 100~TeV
$\gamma$-rays. It is well known that the multi-wavelength
non-thermal energy spectrum is dominated by synchrotron radiation
at energies lower than 1~GeV and by the inverse-Compton
scattering above 1~GeV \cite{Jag96,Ato96}.

TeV $\gamma$-rays from the Crab~Nebula were first clearly detected
by the Whipple collaboration using an imaging air Cherenkov
telescope (IACT) in 1989 \cite{Wee89}.  Since then,
IACT has become the standard telescope for high-energy
$\gamma$-ray observations by virtue of its excellent angular resolution
and efficiency. Many IACTs have been constructed and are operating around
the world, detecting about 70 $\gamma$-ray sources in total up to
the present. On the other hand, air shower arrays have been constructed to
search for $\gamma$-rays from point sources at high altitude. The
merit of this technique is that it can be operated for 24 hours
every day, regardless of weather, with a wide field of view of about
2~sr. The energy threshold of the $\gamma$-rays detected is higher
than that of IACTs, say about 3~TeV at high altitude.  Among these
instruments, the Tibet AS$\gamma$ Collaboration achieved the first successful
observation of the Crab~Nebula at a multi-TeV region in
1999, using a so-called HD (high density) array in which 109 scintillation detectors
were deployed at 7.5~m spacing lattice intervals in an area of
5,175~m$^2$ \cite{Ame99a}.  The Milagro group also reported the
detection of TeV $\gamma$-ray signals from the Crab~Nebula using a
water Cherenkov pool \cite{Atk03}.

Recently, IACTs have obtained updated information on the Crab~Nebula.  The
HEGRA experiment has extended the nebula's energy spectrum
up to 80~TeV with an approximate power-law shape, after patient observation
for almost 400 hours in total \cite{Aha04}.  In contrast, the MAGIC
experiment, equipped with the world's largest tessellated reflector (17~m
in diameter), has successfully observed the lower energy part of the spectrum down to
77~GeV \cite{Alb08}.  The H.E.S.S. group examined the energy spectra
of the Crab~Nebula obtained from various IACTs (Whipple, HEGRA, CAT
and H.E.S.S.) to evaluate the systematic errors of these instruments
\cite{Aha06}.  Their fluxes are well in agreement with one another
within the statistical and systematic errors at the moderate energy
region, although the cutoff energy and spectral index seem to
differ somewhat.  Thus, the Crab~Nebula has been well studied by
various techniques and has been used as a standard calibration source
among the ground-based $\gamma$-ray experiments in the TeV region.

The energy of a primary cosmic-ray particle is estimated by observing
the number of secondary particles in an air shower experiment as well as the
number of Cherenkov photons for IACTs, and then by comparing these values against the results of detailed Monte
Carlo simulations, including the detector structure and response.  The
most conventional method for estimating the absolute energy scale of
primary particles may be to compare the flux values between the direct
(satellite/balloon-borne) and indirect observations. The cosmic-ray
flux, however, depends inevitably on the detection technique,
analysis method, detector simulation, and so on.  Therefore, it is
very important to develop a new method for directly estimating
the absolute energy scale in an air shower experiment at TeV energies.

Clark \cite{Cla57} anticipated in 1957 that the sun and the moon, each with
a finite size of $0\fdg5$ in diameter, cast shadows in the high-energy 
cosmic-ray flux, respectively.  Actually, the shadowing effect of the sun and
the moon (hereafter, we call these the sun's shadow and the moon's
shadow, respectively) was observed by air shower experiments in
the 1990s \cite{Ale91, Ame93}, and the sharpness of the observed
shadows was used to estimate their angular resolutions experimentally.  In
particular, the Tibet air shower array, with its high event trigger rate
and good angular resolution, enables us to use the geomagnetic field as
a magnetic spectrometer for primary cosmic rays at multi-TeV energies.
As almost all primary cosmic rays are positively charged, they are
bent eastward by the geomagnetic field at Yangbajing in Tibet.  The
moon's shadow should then be observed in the
west of the moon's apparent position, although the position of the shadow
in relation to that of the moon depends on the
cosmic-ray energy \cite{Ame00a}. Hence, the position and the
shape of the moon's shadow allow us to estimate the possible
systematic error in the absolute energy scale of observed showers.
Until now, the pointing accuracy and angular resolution of the
Tibet-III array have been checked by monitoring the moon's shadow
continuously month by month \cite{Ame03}.  It is also worthwhile to
note that the moon's and sun's shadows provide information about
the cosmic-ray $\bar{p}/p$ flux ratio \cite{Ach05,Ame07a} and the
global structure of the interplanetary magnetic field between the sun
and the earth \cite{Ame93,Ame94,Ame99b,Ame00a}.

In this paper, we first discuss the systematic uncertainties of the
Tibet-III array and a new method for calibrating the absolute energy
of primary particles in the multi-TeV energy region using the moon's
shadow data.  Based on the results, we report on the $\gamma$-ray
observation of the Crab~Nebula and search for pulsed $\gamma$-ray
emissions from the Crab pulsar using the dataset obtained by the
Tibet-III array.

\section{EXPERIMENT}

\subsection{Tibet-III Air Shower Array}

\begin{figure}
\epsscale{1.10}
\plotone{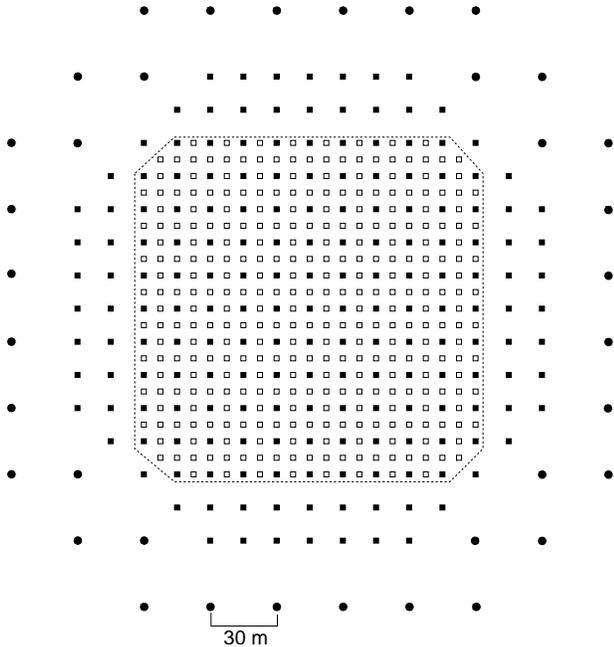}
\caption{ Schematic view of the Tibet-III array operating at
  Yangbajing.  Open squares: FT detectors equipped with a fast-timing (FT)
  photomultiplier tube (PMT); filled squares: FT detectors with a wide
  dynamic-range PMT; filled circles: density detectors with
  a wide-dynamic range PMT. We have selected air shower events whose cores
  are located within the detector matrix enclosed by the dotted line.}
\label{fig1}
\end{figure}

The Tibet-III array shown in Figure~\ref{fig1} was completed on the
basis of the success of the Tibet-I, II and II/HD experiments
\citep{Ame92, Ame99a,Ame00b} in the late fall of 1999.  This array
consists of 533 scintillation detectors of 0.5~m$^{2}$, and the detectors
on the inner side of the array are placed on a lattice with 7.5 m
spacing, covering 22,050~m$^2$.  A lead plate 0.5 cm thick is placed on
top of each detector to improve the angular resolution.  Using this
array, we succeeded in observing $\gamma$-ray flares from Mrk~421 and
found a correlation between TeV $\gamma$-ray and X-ray intensities
\cite{Ame03}.  In 2002 and 2003, the inside area of the
Tibet-III array was further enlarged to 36,900~m$^2$ by adding 256
detectors. This full Tibet-III array has been successfully operating
since 2003.  In this paper, to keep the form of the data the same throughout the
observation period from 1999 to 2005, we reconstructed air shower data
obtained from the detector configuration shown in Figure~\ref{fig1}
even for the full Tibet-III array.

\subsection{Event Reconstruction}

The raw data obtained from the Tibet-III array system mainly consist
of the following: a trigger time stamp for each event from a global positioning system (GPS)
clock supplemented by a computer clock; timing and charge information from
each hit PMT, digitized by a time-to-digital converter
(TDC) and a charge-sensitive analog-to-digital converter (ADC); and
calibration data taken every 20 minutes.  The ADC and TDC counts are then
converted to the number of particles and the relative timing for each
detector, respectively, using the calibration data. An air
shower event is reconstructed as follows.

The core position of each air shower is estimated using the lateral
distribution of the number of shower particles observed in the
array. The density-weighted position of the air shower core on the surface
of the Tibet-III array is calculated as $(X_{\rm core}, Y_{\rm
  core})~=~\left(\frac{\sum_{i}~\rho_{i}^{2}~x_{i}}{\sum_{i}~\rho_{i}^{2}},
\frac{\sum_{i}~\rho_{i}^{2}~y_{i}}{\sum_{i}~\rho_{i}^{2}}\right)$,
where $x_{i}$ and $y_{i}$ are the coordinates of the {\it i}-th detector and
$\rho_{i}$ is the number density (m$^{-2}$) of detected particles.  In
this analysis, we regard the shower size $\sum\rho_{\rm FT}$ as the
primary energy reference, where size $\sum\rho_{\rm FT}$ is
defined as the sum of the number of particles per m$^2$ for each FT
detector.  For $\gamma$-ray-induced air showers, overall core position resolutions
are then estimated as 12~m and 4~m at median for $\sum\rho_{\rm FT} < 100$ and
$\sum\rho_{\rm FT} > 100$, respectively.

The arrival direction of each shower is estimated assuming that the front of
the air shower is conical shape.  The apex of a cone is then taken to be
the estimated core position $(X_{\rm core}, Y_{\rm core})$.  The
average delay time $T$ (ns) of shower particles is expressed as a
function of the distance $R$ (m) from the core position as $T = 0.075
R$, which is optimized by simulations.  This gives a
cone slope of $1\fdg3$ with respect to the plane perpendicular to the
air shower arrival direction.

\subsection{Event Selection} \label{s-2.3}

An event trigger signal is issued when an any-fourfold coincidence
appears in the FT detectors that have each recorded more than 0.6 particles within a
coincidence gate width of 600~ns, resulting in a trigger rate of
about 680~Hz.  We collected 2.0$\times$10$^{10}$ events during 1318.9
live days from 1999 November 18 through 2005 November 15 after some
quality cuts and event selection based on three simple
criteria: (1) each shower event should fire four or more FT detectors
that have each recorded 1.25 or more particles; (2) among the 9 hottest FT detectors
in each event, 8 should be contained in the fiducial area enclosed by
the dotted line in Figure~\ref{fig1}.  If fewer than 9 detectors have been hit,
they should all be contained in the fiducial area;
and (3) the zenith angle of the event arrival direction should be less
than 40$\degr$.  After these criteria have been met, the
overall angular resolution and the modal energy of air shower events,
thus obtained, are better than 1~degree and about 3~TeV, respectively
\cite{Ame03}, thereby covering the upper portion of the energies measured by
IACTs.

\section{MOON'S SHADOW AND PERFORMANCE OF THE TIBET-III AIR SHOWER ARRAY}

\subsection{Analysis} \label{s-3.1}

Using the dataset described in $\S$\ref{s-2.3}, we further select the
events within a circle of the radius 5$\degr$ centered at the moon;
this circle is defined as the on-source field. An equatorial coordinate
system is defined, fixing the origin of coordinates at the moon's
center.  To estimate the background against deficits in the moon's
shadow, we adopt the equi-zenith angle method \cite{Ame03,Ame05}, which
is also used for the Crab~Nebula observation described in
$\S$\ref{s-4.1}.  Eight off-source fields are symmetrically aligned on
both sides of the on-source field, at the same zenith angle. In order
to avoid deficit events that are affected by background event contamination 
of the on-source field, the nearest two off-source fields
are each set at an angular distance $9\fdg6$ from the on-source field.  Other
off-source fields are located every $3\fdg2$ from the nearest
off-source fields.  The position of each observed event in
an on-/off-source field is then specified by the angular distance
$\theta$ and the position angle $\phi$, where $\theta$ and $\phi$ are
measured from the center and from the north direction, respectively.
Using $\theta$ and $\phi$, the on-/off-source fields are meshed by
$0\fdg05$$\times$$0\fdg05$ cells, and we count the number of events in
each cell.  To maximize the $S/N$ ratio, we group the cells into new
on-/off-source bins according to the angular resolution, which depends
on the shower size $\sum\rho_{\rm FT}$.  The angular resolution
becomes worst at a threshold energy of around 1~TeV. This value is,
however, smaller than half of the angular distance between two
adjoining off-source bins, i.e., $\sim1\fdg6$, so that off-source
bins never overlap mutually in this background analysis.

We calculate the statistical significance of deficits or signals using
the formula \cite{Li83} $(N_{\rm ON} - \epsilon N_{\rm
  OFF}) / \sqrt{\epsilon (N_{\rm ON} + N_{\rm OFF})}$, where $N_{\rm
  ON}$ and $N_{\rm OFF}$ are the number of events in the
on-source bin and the number of background events summed over 8
off-source bins, respectively, and  $\epsilon$ is the ratio of the on-source solid angle area to
the off-source solid angle area ($\epsilon$ = 1/8 in this work).

In order to investigate the energy dependence, the shower size
$\sum\rho_{\rm FT}$ is divided by 1/4 decades in 10$<$$\sum\rho_{\rm
  FT}$ $\leq$100, and by 1/3 decades in 100$<$$\sum\rho_{\rm
  FT}$$\leq$1000, where the lowest air shower size bin
is omitted from the analysis, because it is close to the energy
threshold of the Tibet-III array and the trigger efficiency is
estimated to be very low ($<$1\%).  Hereafter, this partition is
commonly used in observations of both the moon's shadow and the
Crab~Nebula.

Figure~\ref{fig2} shows the experimental significance map of the deficit event
densities observed with the Tibet-III array for 1318.9 live days.
This map is smoothed using the events with $\sum\rho_{\rm
  FT}$$>$10$^{1.25}$ ($>\sim$2~TeV) within a circle of radius
$0\fdg9$, corresponding to the overall angular resolution for these
events.  The maximum deficit reaches the significance level of
45~$\sigma$ at the center.  It is seen that the center of the observed
moon's shadow is shifted to the west by about $0\fdg2$ due to the
effect of the geomagnetic field.

\subsection{Monte Carlo Simulation of the Moon's Shadow} \label{s-3.2}

\begin{figure}[t]
\epsscale{1.20}
\plotone{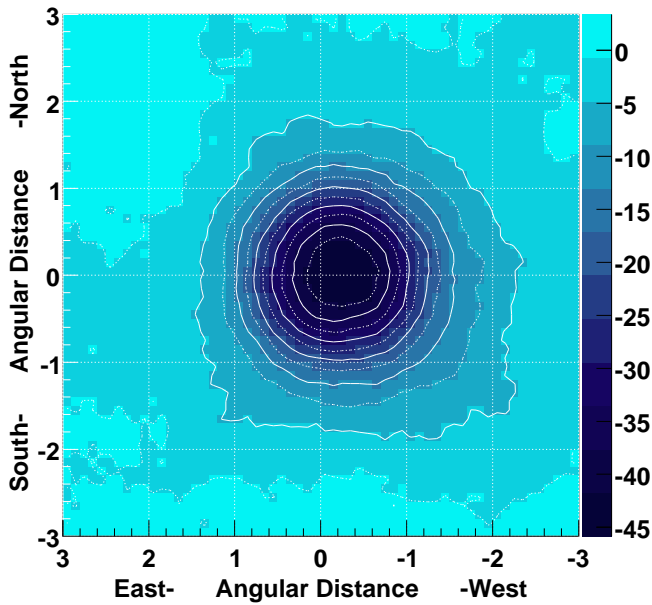}
\caption{ Significance map of the deficit event densities observed by
  the Tibet-III array for 1318.9 live days, made using the events with
  $\sum\rho_{\rm FT} >10^{1.25}$ ($>\sim$2~TeV), in the square area of
  6$\degr$$\times$6$\degr$ whose origin is at the apparent center of the moon.
  The scale at right shows the level of significance
  of the deficit event density in terms of the standard deviation
  $\sigma$.}
\label{fig2}
\end{figure}

\begin{figure}[t]
\epsscale{1.20}
\plotone{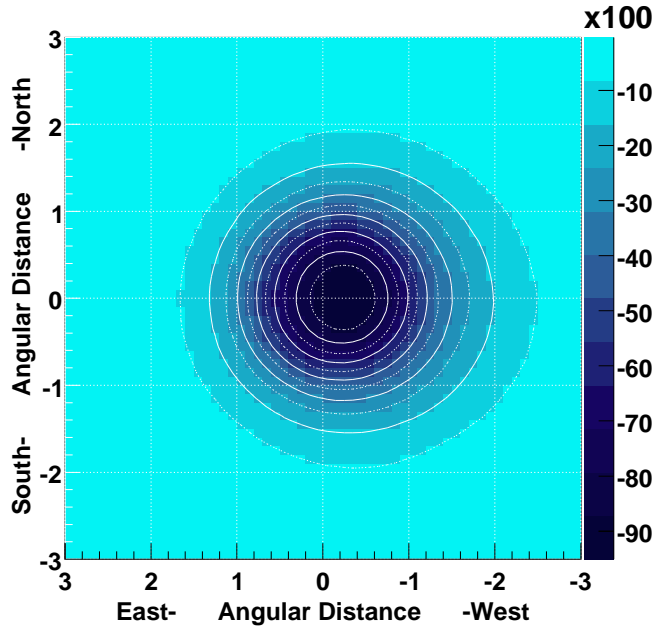}
\caption{ Deficit event density map obtained by the MC simulation.
  The events with $\sum\rho_{\rm FT} >10^{1.25}$ ($>\sim$2~TeV) are
  plotted in the square area of 6$\degr$$\times$6$\degr$.  The map is
  made in the same way as in Figure~\ref{fig2}, and the scale at right
  represents the deficit event density (degree$^{-2}$).}
\label{fig3}
\end{figure}

We have performed a detailed Monte Carlo (MC) simulation of the moon's
shadow.  For the geomagnetic field, we adopt the International
Geomagnetic Reference Field (IGRF) 9th generation model \cite{Mac03}
at an altitude $<$~600~km, and connect it to the dipole moment model at
an altitude $>$ 600~km.  For the primary particles, we use the chemical
composition obtained mainly from the data of direct observations
\cite{Asa98,San00,Apa01,Ame08} in the energy range of 0.3~TeV to
1000~TeV.  The minimum energy of primary particles is set to 0.3~TeV,
which is low enough to cover the threshold energy of our trigger
condition.  Air shower events are generated at the top of the
atmosphere along the moon's orbit around the earth, using the CORSIKA
code \cite{Cor98} with QGSJET or SIBYLL interaction models.  The air
shower core of each simulated event is uniformly distributed over a
circular region with a 300~m radius centered at the array; this circle
sufficiently covers the area where cosmic-ray events are actually
triggered in our array.  Air shower particles generated by primary
particles in the atmosphere are traced until their energies reach
1~MeV.  In order to treat the MC events in the same way as the events in the
experimental data, these simulated events are distributed among the detectors
in the same detector configuration as in the Tibet-III array by the
Epics code \cite{Ame08,Kas}, and are converted to the same format as
the experimental dataset, such as the ADC and TDC values at each
detector.  After air shower reconstruction analysis and data
selection, we assign the opposite charge to the remaining primary
particles.  These anti-particles are randomly shot back toward
directions within 20$\degr$$\times$20$\degr$ centered at the moon from
the first interaction point of the air shower. Hereafter, we call
this the initial shooting direction. The particle track
influenced by the geomagnetic field between the earth and the moon is
calculated by the Runge-Kutta method of order 4 based on the Lorentz
force.  If the primary particle hits the moon, its initial shooting
direction should be equivalent to the observed particle direction
shielded by the moon.  Otherwise, it is shot back in another
direction and the particle track is calculated again. This routine
continues until almost all particles have hit the moon.  Finally, these
initial shooting directions are smeared by the angular resolution
event by event.  In this way, the expected moon's shadow
is equivalent to the observed moon's shadow.  Figure~\ref{fig3}
shows the event map of the moon's shadow calculated by the MC
simulation for events with $\sum\rho_{\rm FT}$$>$10$^{1.25}$
($>\sim$3~TeV).  This map is smoothed using the events within a circle
of radius $0\fdg9$, corresponding to the overall angular resolution
for events with $\sum\rho_{\rm FT}$$>$10$^{1.25}$.  The MC simulation
well reproduces the observed moon's shadow, as shown in
Figure~\ref{fig2}.

\subsection{Shape of the Moon's Shadow and Performance of the Tibet-III Array} \label{s-3.3}

\begin{figure*}[!t]
\epsscale{1.10}
\plotone{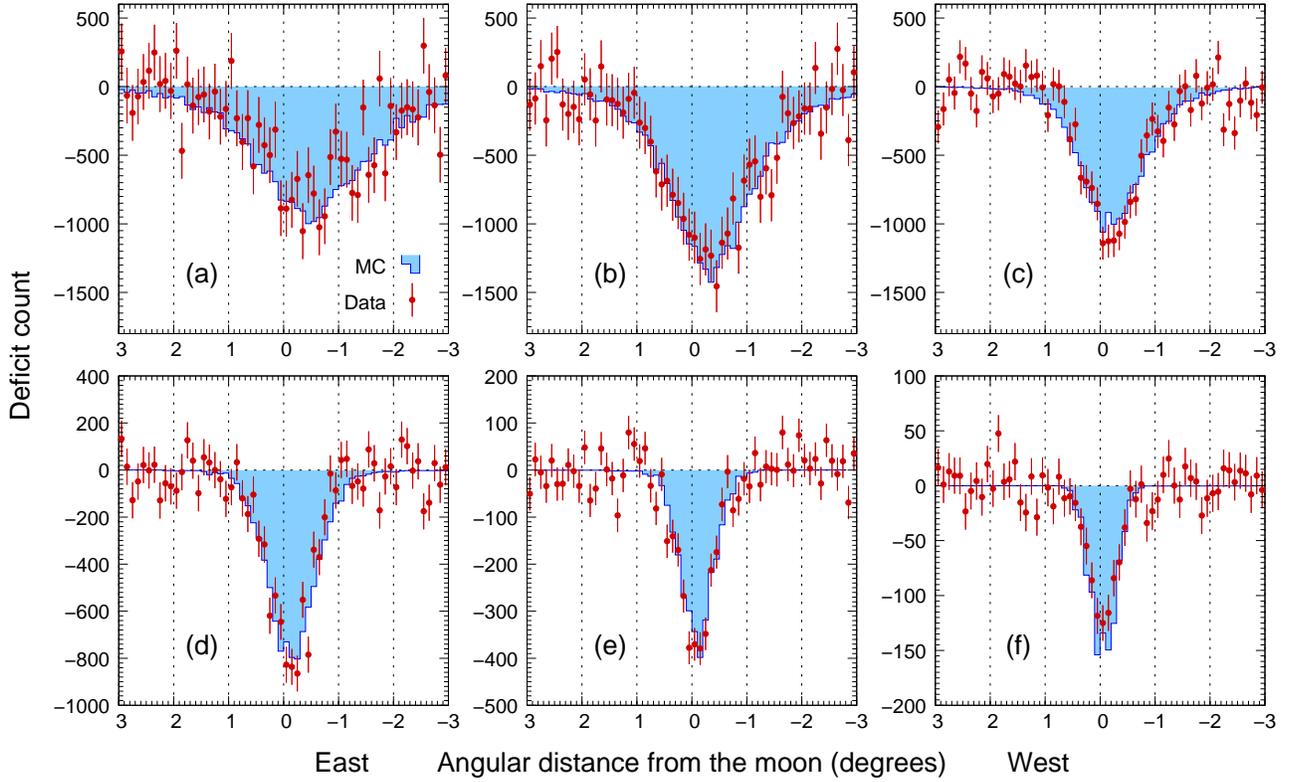}
\caption{Filled circles show experimental data for deficit counts
  around the moon projected to the east-west axis for each
  $\sum\rho_{\rm FT}$ bin.  We use the events contained in the angular
  band, centered at and parallel to the east-west axis, comparable to
  the $\sum\rho_{\rm FT}$-dependent angular resolution:
(a): $\pm1\fdg4$ for 2.94~TeV/$Z$ in $10^{1.25} < \sum\rho_{\rm FT} \le 10^{1.50}$;
(b): $\pm1\fdg0$ for 4.20~TeV/$Z$ in $10^{1.50} < \sum\rho_{\rm FT} \le 10^{1.75}$;
(c): $\pm0\fdg7$ for 6.46~TeV/$Z$ in $10^{1.75} < \sum\rho_{\rm FT} \le 10^{2.00}$;
(d): $\pm0\fdg5$ for 11.4~TeV/$Z$ in $10^{2.00} < \sum\rho_{\rm FT} \le 10^{2.33}$;
(e): $\pm0\fdg3$ for 21.6~TeV/$Z$ in $10^{2.33} < \sum\rho_{\rm FT} \le 10^{2.67}$;
(f): $\pm0\fdg2$ for 45.4~TeV/$Z$ in $10^{2.67} < \sum\rho_{\rm FT} \le 10^{3.00}$.
The solid histograms denote the moon's shadow simulation
assuming the primary cosmic-ray composition based on direct
observations \cite{Asa98,San00,Apa01,Ame08}.}
\label{fig4}
\end{figure*}

The filled circles in Figure~\ref{fig4}~(a)--(f) show the observed
deficit counts around the moon projected onto the east-west axis for
each $\sum\rho_{\rm FT}$ bin, where $\sum\rho_{\rm FT}$ is the shower
size defined as the sum of the number of particles per m$^2$ for each
FT detector. The representative cosmic-ray energy in each
$\sum\rho_{\rm FT}$ bin is defined as be the logarithmic mean of the
energy $E$ divided by the atomic number $Z$, assuming the cosmic-ray
composition spectrum described in $\S$\ref{s-3.2}.  In these figures,
it is seen that the peak position of the deficit counts gradually shifts
to the west as primary energy decreases due to the
influence of the geomagnetic field.  Also, the deficit counts
become narrower as primary energy increases, since the angular
resolution increases roughly in inverse proportion to
$\sqrt{\sum\rho_{\rm FT}}$.  The MC results (histograms) are compared
with the experimental data at various size intervals in
Figure~\ref{fig4}, and are in good agreement with the experimental data.

In order to estimate the peak position of the observed moon's shadow,
we use the shadow shape obtained by the MC simulation.
We first express the MC shadows shown in Figure~\ref{fig4} with
the superposition of two Gaussian functions using the least $\chi^{2}$
method as,
\begin{eqnarray}
\nonumber
f_{\rm MC}(\theta) = G_{1}(\theta; a_1, m_1, \sigma_1) \\
+ G_{2}(\theta; a_2, m_2, \sigma_2),
\label{equ1}
\end{eqnarray}
where $G_{i}(\theta; a_i, m_i, \sigma_i) = a_{i}e^{
  -(\theta-m_{i})^2/\sigma_{i}^2 }$ and $\theta$ is the angular
distance from the moon.  Here, $a_1$, $a_2$, $m_1$, $m_2$, $\sigma_1$ and
$\sigma_2$ are the fitting parameters denoting the amplitudes, means
and one-standard deviations of the double Gaussian, respectively.  It
is found that the shadows (e) and (f) can be expressed by a single
Gaussian, while the others are expressed by a double Gaussian.

Using these fitting parameters, we then estimate the peak position of
the observed shadow as follows. Keeping the form of function
$f_{\rm MC}(\theta)$, we express the observed shadow by the
equation,
\begin{eqnarray}
\nonumber
f_{\rm Data}(\theta)  =  G_{1}(\theta;A_1, M_1, \sigma_1) \\
+ G_{2}(\theta; A_1\times(a_2/a_1), M_1+(m_2-m_1), \sigma_2), &
\label{equ2}
\end{eqnarray}
where $A_1$ and $M_1$ denote the amplitude and mean of the Gaussian,
respectively, and are free parameters, while the others are the
coefficients calculated by fitting equation~(\ref{equ1}).  Using
equation~(\ref{equ2}), we can estimate the peak position of the moon's
shadow.

\begin{figure}
\epsscale{1.20}
\plotone{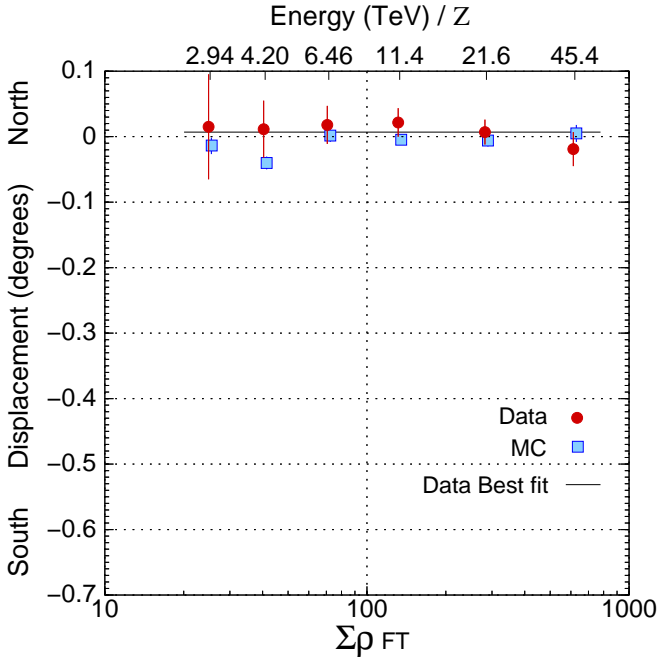}
\caption{Dependence of shower size on the displacement of the moon's
  shadow in the north-south direction.  The filled circles and open
  squares represent experimental data and the MC simulation,
  respectively.  The solid line denotes the fitting to the
  experimental data assuming a constant function, resulting in
  $0\fdg008\pm0\fdg011$. The upper scale indicates the logarithmic mean
  of $E/Z$ (TeV/$Z$) in each $\sum\rho_{\rm FT}$ bin. }
\label{fig5}
\end{figure}

We also confirm that the east-west component of the geomagnetic field strength
is negligible in the part of the sky where the moon is visible by
the Tibet-III array. This means that the north-south displacement of the moon's shadow
observed by the Tibet-III array does not depend on the geomagnetic field.
 The displacement of the peak of the moon's shadow in the north-south 
direction then enables us to estimate
the magnitude of the array's systematic pointing error.  Figure~\ref{fig5}
shows the energy dependence of the displacement of the moon's shadow
in the north-south direction.  The filled circles denote the
experimental data, and the open squares are the MC results. 
The MC simulation well reproduces the experimental data.  A
$\chi^{2}$ fitting to the data gives $0\fdg008\pm0\fdg011$ assuming a
constant function independent of energy.  From this, the systematic
pointing error is estimated to be smaller than $0\fdg011$.

\subsection{Calibration of Primary Particle Energy and Systematic Errors} \label{s-3.4}

\begin{figure}
\epsscale{1.20}
\plotone{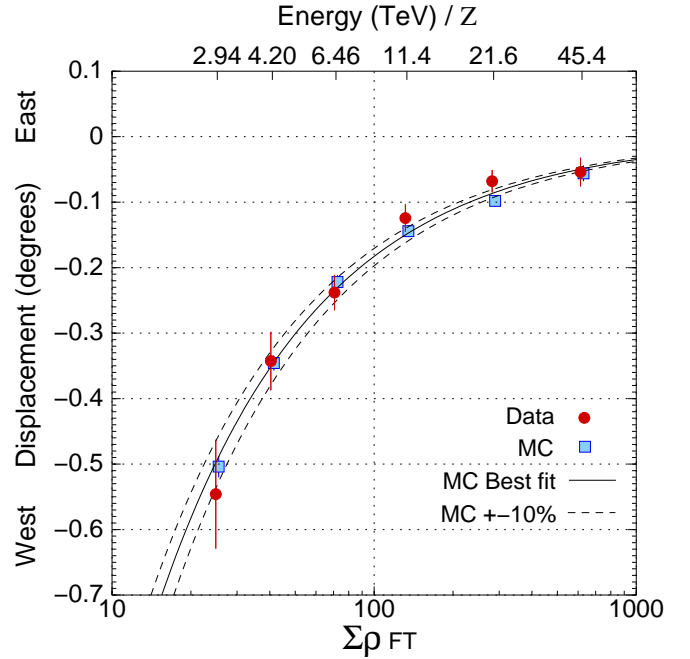}
\caption{Dependence of shower size on the displacement of the moon's
  shadow in the east-west direction. The filled circles show the
  experimental data, and open squares represent the MC
  simulation. The solid curve is fitted to the MC events, and dashed
  curves show a $\pm$10\% deviation from the solid curve,
  respectively.  The upper scale indicates the logarithmic mean of $E/Z$
  (TeV / $Z$) in each $\sum\rho_{\rm FT}$ bin. }
\label{fig6}
\end{figure}

Figure~\ref{fig6} shows the shower size dependence of the displacement
of the moon's shadow in the east-west direction, obtained by fitting Figure~\ref{fig4}. In this figure, the
open squares show the MC results using the QGSJET model, and are quite
consistent with the experimental data. The upper scale indicates the
logarithmic mean of the energy $E$ divided by the atomic number $Z$
(TeV$/$$Z$), i.e., $<\log(E/Z)>$, in each $\sum\rho_{\rm FT}$ bin.
One can see that the position of the moon's shadow gradually shifts to
the west as the primary energy decreases due to the influence
of the geomagnetic field.  Hence, the absolute energy scale of cosmic
rays observed by the Tibet-III array can be directly checked by using
the geomagnetic field as a magnetic spectrometer, as we now discuss.

First, the MC simulation points are fitted by the function
$\kappa(\sum\rho_{\rm FT}/100)^{\lambda}$ to define a standard
curvature function, resulting in $\kappa = -0.183$ and $\lambda = -
0.720$, as shown by a solid curve in Figure~\ref{fig6}, where the MC
statistical errors are negligible compared with the experimental data.

Second, the experimental data (filled circles) are fitted by this
standard curvature function with the $\sum\rho_{\rm FT}$ shift term
\begin{equation}
-0.183~[(1 - \Delta R_{\rm S})(\sum\rho_{\rm FT}/100)]^{-0.720},
\label{equ3}
\end{equation}
to estimate the possible shift in the $\sum\rho_{\rm FT}$ between the
experimental data and the MC simulation, as shown by the solid curve in
Figure~\ref{fig6}, where $\Delta$$R_{\rm S}$ is the $\sum\rho_{\rm
  FT}$ shift ratio, resulting in $\Delta$$R_{\rm S} = (-4.9 \pm
9.5)\%$.  We should then convert $\Delta$$R_{\rm S}$ to the energy
shift ratio $\Delta$$R_{\rm E}$ as a final result.  To determine the
relationship between $\Delta$$R_{\rm S}$ and $\Delta$$R_{\rm E}$, and to
confirm that this method is sensitive to energies, we prepare six
kinds of MC event samples in which the energy of the primary particles is
systematically shifted event by event in the moon's shadow
simulation. These six $\Delta$$R_{\rm E}$s are $\pm$20\%, $\pm$15\%
and $\pm$8\%, respectively.  In each MC event sample, the $\sum\rho_{\rm
  FT}$ dependence of the displacement of the moon's shadow is
calculated in the same way, and the $\sum\rho_{\rm FT}$ shift ratio
$\Delta$$R_{\rm S}$ is estimated by fitting the data to
equation~(\ref{equ3}).  Finally, we get the relation $\Delta$$R_{\rm E}$
= $(-0.91\pm0.05)$ $\Delta$$R_{\rm S}$ assuming a linear function.
Hence, the systematic error in the absolute energy scale
$\Delta$$R_{\rm E}$ with statistical error $\sigma_{\rm stat}$ is
estimated to be $(+4.5 \pm 8.6_{\rm stat})\%$.

\begin{figure}[t]
\epsscale{1.20}
\plotone{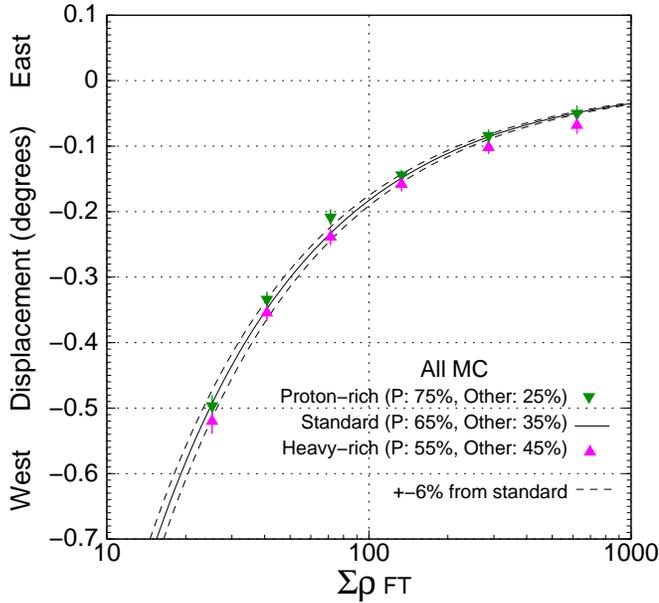}
\caption{Dependence of shower size on the displacement of the moon's
  shadow in the east-west direction by the MC simulation for the
  different primary composition models. The filled circles and open
  squares show the experimental data and MC simulation, respectively.  The
  solid curve denotes the best-fit curve for the same standard
  composition ratio as in Fig.~\ref{fig6} (65\% P after triggering by
  the Tibet-III array, where P means protons). The dashed curves show
  a 6\% shift, corresponding to $\sigma_{\rm syst1}$, from the solid
  curve (see text).  The downward and upward triangles are the
  simulated results for the proton-rich model (P:75\%), and for the
  heavy-rich model (P:55\%), respectively. }
\label{fig7}
\end{figure}

\begin{figure}[t]
\epsscale{1.20}
\plotone{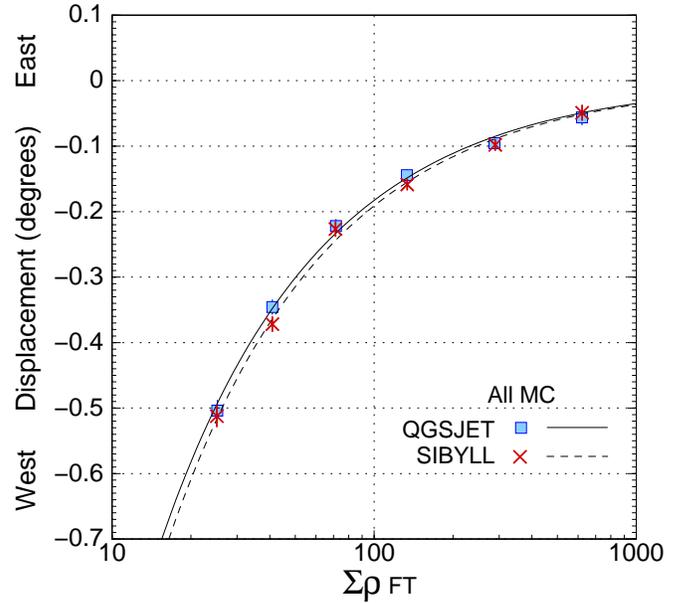}
\caption{Dependence of shower size on the displacement of the moon's
  shadow in the east-west direction by the MC simulation for different
  hadronic interaction models.  The open squares and cross marks are
  the results obtained by the QGSJET and SIBYLL models, respectively.  The solid
  and dashed curves are the best-fit results assuming the QGSJET and
  SIBYLL models, respectively.}
\label{fig8}
\end{figure}

Furthermore, we investigate two kinds of systematic uncertainties with
the proposed method.  One is that the position of the moon's shadow by the MC
simulation depends on the assumed primary cosmic-ray composition.  In
this simulation, the chemical composition ratio of primary cosmic rays
is estimated based mainly on the data obtained by direct observations.
These datasets should also have statistical and systematic errors.
The position of the moon's shadow is dominated by the light component,
so that the proton ratio is artificially varied by $\pm$10\% from a
standard chemical composition without changing their spectral index,
while the other components heavier than helium are varied by $\mp$10\%
in total.  Figure~\ref{fig7} shows the results for the composition
dependence of primary cosmic rays.  The downward triangles are the
results obtained by the proton-rich model (75\% protons after triggering by the
Tibet-III array), while the upward triangles are the ones for the
heavy-rich model (P:55\%).  These models are fitted by
equation~(\ref{equ3}). We then obtain $\sigma_{\rm syst1}$ = $\pm$6\%
for the systematic error due to the difference in chemical composition,
as shown by the dashed curves in Figure~\ref{fig7}.  Another
systematic uncertainty is caused by the difference between hadronic
interaction models.  Figure~\ref{fig8} compares the results for the
hadronic interaction model dependence by QGSJET with those obtained by SIBYLL.  It
is found that the results by the SIBYLL model can be well fitted by
equation~(\ref{equ3}) obtained using the QGSJET model.  We then
obtain $\sigma_{\rm syst2}$ = 6\% difference between two models.
Finally, the difference in the energy dependence of the moon's shadow
between the experimental data and the MC events is estimated to be
$+4.5\%~(\pm 8.6_{\rm stat} \pm 6_{\rm syst1} \pm 6/2_{\rm syst2})\%$.
This value is within the statistical and systematic errors.  Hence,
the absolute energy scale error in the Tibet-III array is estimated to
be smaller than 12\% = $\sqrt{\Delta R_{\rm E}^2 + \sigma_{\rm stat}^2
  + \sigma_{\rm syst1}^2 + (\sigma_{\rm syst2}/2)^2}$ in total
averaged from 3 to 45 (TeV$/$$Z$).

\subsection{On the Energy Estimation of $\gamma$-Ray Showers} \label{s-3.5}

We established a new calibration method of the absolute energy scale
for cosmic rays based on the moon's shadow analysis as described
above.  The air shower induced by the primary cosmic ray consists of
high-energy hadronic and electromagnetic cascades. Although several
plausible hadronic interaction models are prepared in the MC
simulation, there still remain dependence between these
models. Therefore, the energy reconstruction from the air-shower size
depends on hadronic interaction models and also the primary chemical
composition models. These systematic errors were adequately taken into
account in the absolute energy scale error in this paper.  On the
other hand, the air shower induced by the primary $\gamma$-ray is
predominated by the theoretically well-known electromagnetic cascades,
because the photon cross section for hadronic interactions is
approximately two orders of magnitude smaller than that for the pair
creation process. Hence, we naturally expect that the absolute energy
scale error for $\gamma$-rays is smaller than 12\% which is deduced
from the moon's shadow observed in the cosmic rays described in
$\S$\ref{s-3.4}. In the next section, we will provide reliable results
on the multi-TeV $\gamma$-ray observation from the Crab~Nebula with
the Tibet-III air shower array finely tuned by the cosmic-ray moon's
shadow.

\section{MULTI-TeV $\gamma$-RAY OBSERVATION FROM THE CRAB}

\subsection{Analysis} \label{s-4.1}

\begin{figure}
\epsscale{1.20}
\plotone{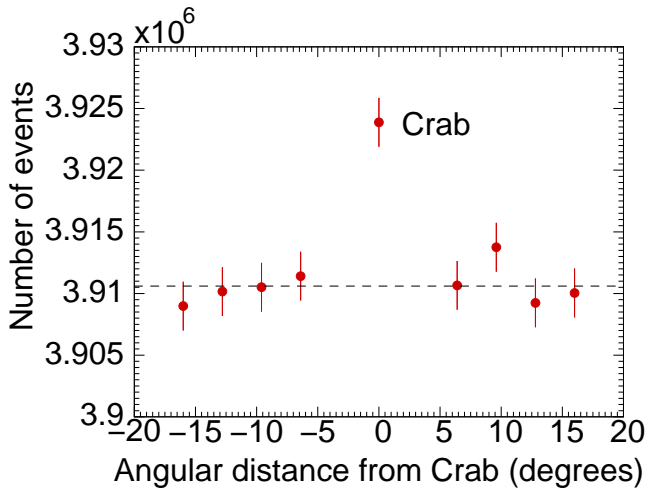}
\caption{Number of observed air shower events with $\sum\rho_{\rm FT}
  >10^{1.25}$ ($>$ $\sim$1~TeV) after event reduction for the
  observation time of 1318.9 detector live days as a function of
  angular distance from the Crab~Nebula in the azimuthal direction. }
\label{fig9}
\end{figure}

\begin{figure}
\epsscale{1.20}
\plotone{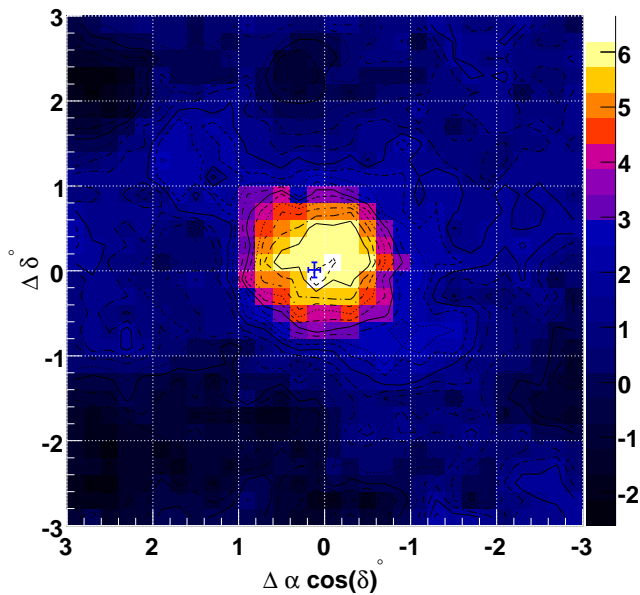}
\caption{Contour map of significance distribution around the Crab~Nebula 
  ($\alpha =83\fdg63$, $\delta = 22\fdg02$) for events with
  $\sum\rho_{\rm FT} >10^{1.25}$ ($>$ $\sim$1~TeV). A clear peak
  excess is seen at the center position
  $\Delta\alpha$cos($\Delta\delta$) = $\Delta\delta$ = 0\degr, where
  $\Delta\alpha$ and $\Delta\delta$ are the relative right accension and
  declination, respectively, from the Crab~Nebula.
  The cross mark indicates the pointing error by a point spread function fitting.}
\label{fig10}
\end{figure}

In order to extract an excess of TeV $\gamma$-ray events coming from
the direction of the Crab~Nebula ($\alpha = 83\fdg63$, $\delta =
22\fdg02$), we adopt the same method as used in the Mrk~421 analysis in
our previous work \cite{Ame03}.  We call it the equi-zenith angle
method, which is used also for the moon's shadow analysis described in
$\S$\ref{s-3.1}.  The background is estimated by the number of events
averaged over eight off-source bins with the same angular radius as
the on-source bin, at the same zenith angle, recorded at the same time
intervals as the on-source bin. The nearest two off-source bins are
set at an angular distance $6\fdg4$ from the on-source bin to avoid a
possible signal tail leaking into these off-source bins.  Other
off-source bins are located every $3\fdg2$ step from the nearest
off-source bins.  The search window radius is expressed by $6.9/\sqrt{
  \sum\rho_{\rm FT} }$ degrees as a function of $\sum\rho_{\rm FT}$,
which is shown to maximize the $S/N$ ratio by MC study of
$\gamma$-ray observation \cite{Ame03}.

The number of events after the event reduction is plotted in
Figure~\ref{fig9} as a function of angular distance from the
Crab~Nebula in the azimuthal direction. A clear peak of $\gamma$-ray
signals from the Crab~Nebula is seen at 6.3~$\sigma$ statistical
significance above the flat cosmic-ray background.  Figure~\ref{fig10}
is a significance map around the Crab~Nebula. The peak excess is seen
at the Crab~Nebula position.  The pointing error as shown by a cross
mark in Figure~\ref{fig10} is estimated to be
$\Delta\alpha=+0.13\pm0.08$ and $\Delta\delta=+0.01\pm0.09$ by the
point spread function fitting.  This is consistent with the Crab's
position within statistical error.  As difference between $\gamma$-ray
induced air showers and cosmic-ray induced ones does not affect the
pointing accuracy essentially, we estimate our pointing accuracy to
to be $0\fdg011$ deduced from the moon's shadow analysis described in 
$\S$\ref{s-3.3}.

\subsection{Monte Carlo Simulation of $\gamma$-Ray Observation from the Crab~Nebula}

\begin{figure}
\epsscale{1.20}
\plotone{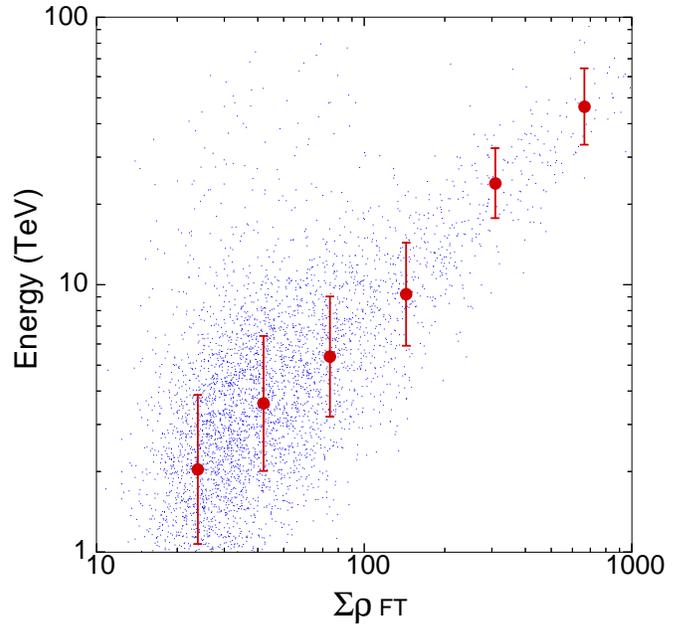}
\caption{Scatter plot of the shower size $\sum\rho_{\rm FT}$
  and the energy of primary $\gamma$-rays, where a differential
  power-law spectrum of the form $E^{-2.6}$ starting at 0.3~TeV is
  assumed for primary $\gamma$-rays. For details, see text.}
\label{fig11}
\end{figure}

\begin{figure}
\epsscale{1.18}
\plotone{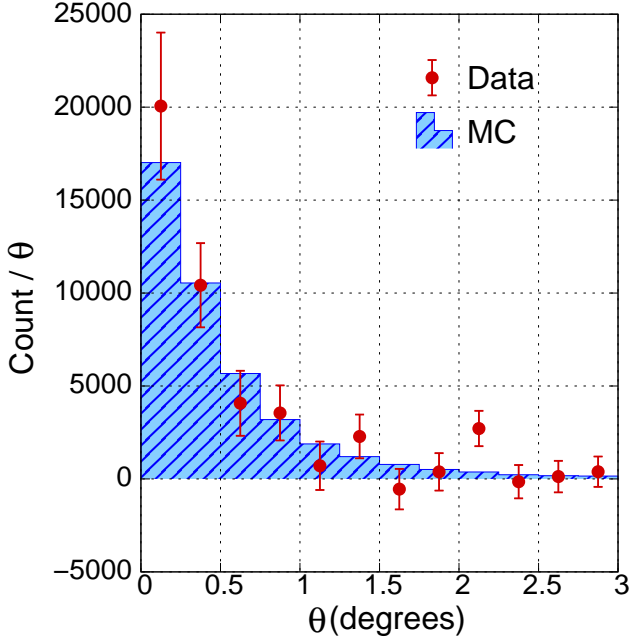}
\caption{Distribution of excesses as a function of the opening angle
  relative to the Crab~Nebula direction $\theta$. The filled circles
  and shaded histograms stand for the experimental data and the MC
  events with $\sum\rho_{\rm FT} >10^{1.25}$ ($>$ $\sim$1~TeV),
  respectively.}
\label{fig12}
\end{figure}

The performance of the Tibet-III array has been studied by a full MC
simulation using the CORSIKA code \cite{Cor98} for event generation in
the atmosphere and the Epics code \cite{Kas} for the response of
the scintillation detector \cite{Ame03}.  These procedures
are essentially the same as in the case for the moon's shadow described
in $\S$\ref{s-3.2}.  In the simulation for $\gamma$-ray observation from the Crab~Nebula, primary
$\gamma$-rays, assuming the energy spectrum of a power-law type in the
energy region of 0.3~TeV to 1000~TeV, are thrown along the diurnal
motion of the Crab~Nebula in the sky.  The air shower events generated are
uniformly distributed over circle with a 300~m radius centered at the
Tibet-III array.  Shown in Figure~\ref{fig11} is the scatter plot of
shower size $\sum\rho_{\rm FT}$ and the energy of $\gamma$-rays coming
from the Crab direction.  The filled circles and error bars stand for
the logarithmic mean of the energy and one-standard deviation of the 
logarithmic Gaussian, respectively.
The one-event energy resolution is estimated to be approximately
$(-40/+70)\%$ at 10~TeV, and approximately $\pm$100\% in the region of
a few TeV.

The Crab~Nebula can be treated as a point-like source at the TeV energy
region.  To investigate the point spread function of the Tibet-III array, we
compared the $\theta$ distribution of the Crab~Nebula between the
experimental data and the MC events, where $\theta$ is the opening angle
relative to the Crab~Nebula direction.  Figure~\ref{fig12} shows the
distribution of the excess events as a function of $\theta$ for events
with $\sum\rho_{\rm FT} >10^{1.25}$. The experimental data agree well
with the MC simulation assuming the point-like source.

\subsection{Energy Spectrum of $\gamma$-Rays from the Crab Nebula}

\begin{figure}[t]
\epsscale{1.21}
\plotone{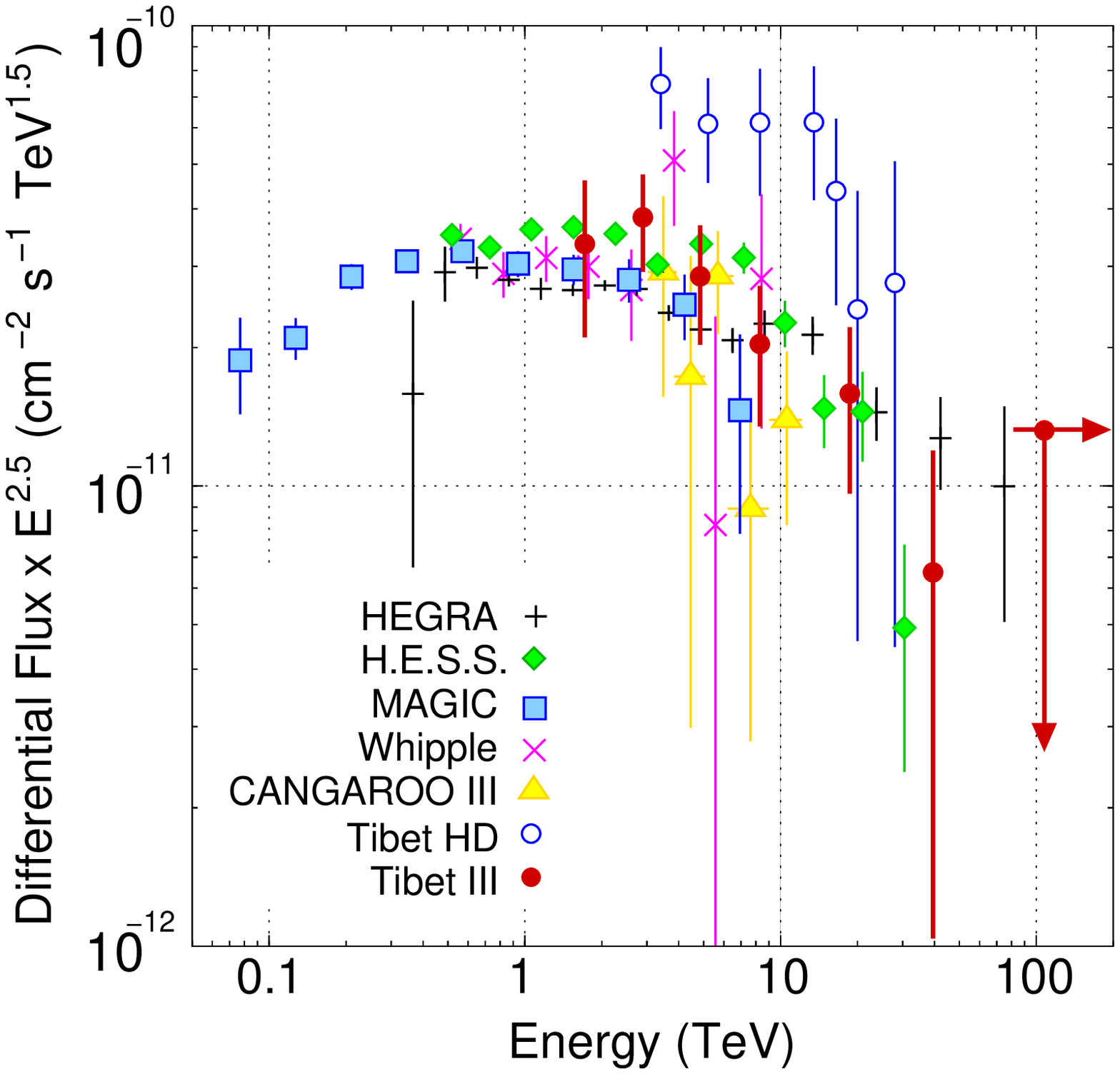}
\caption{Differential energy spectrum of $\gamma$-rays from the
  Crab~Nebula obtained using the data collected from 1999 November to
  2005 November with the Tibet-III array in comparison with the
  results from IACTs: Whipple \cite{Hil98}, HEGRA \cite{Aha04},
  CANGAROO~III \cite{Eno06}, H.E.S.S. \cite{Aha06} and MAGIC \cite{Alb08}.
  The Tibet-III upper limit is given at the 90\%
  confidence level, according to a statistical prescription
  \cite{Hel83}.  }
\label{fig13}
\end{figure}





\begin{table*}
\begin{center}
\caption{Logarithmic mean of energy and differential flux for each
  $\sum \rho_{\rm FT}$ bin as shown in Fig.~\ref{fig13}.
\label{tbl-1}}
\begin{tabular}{ccccc}
\tableline\tableline
$\sum \rho_{\rm FT}$ & Energy &  $N_{\rm ON}$ & $\epsilon N_{\rm OFF}$ & Differential Flux \\
                     & [TeV]  &               &                        & [cm$^{-2}$ s$^{-1}$ TeV$^{-1}$] \\
\tableline
$10^{1.25}$ -- $10^{1.50}$ & 1.71 & 1935499 & 1931547 & $(8.72\pm3.25) \times 10^{-12}$\\
$10^{1.50}$ -- $10^{1.75}$ & 2.89 & 1382356 & 1377139 & $(2.70\pm0.643) \times 10^{-12}$\\
$10^{1.75}$ -- $10^{2.00}$ & 4.84 & 444504 & 442074 & $(5.52\pm1.60) \times 10^{-13}$\\
$10^{2.00}$ -- $10^{2.33}$ & 8.29 & 134509 & 133362 & $(1.03\pm0.348) \times 10^{-13}$\\
$10^{2.33}$ -- $10^{2.67}$ & 18.6 & 21530 & 21138 & $(1.06\pm0.417) \times 10^{-14}$\\
$10^{2.67}$ -- $10^{3.00}$ & 39.5 & 3923 & 3844 & $(6.64\pm5.57) \times 10^{-16}$\\
$>$ $10^{3.00}$            & 107  & 1558 & 1569 & $<1.10 \times 10^{-16}$\\
\tableline
\end{tabular}
\end{center}
\end{table*}



The $\gamma$-ray flux from the Crab~Nebula is estimated by assuming a
power-law spectrum $f(E) = \alpha E^{\beta}$.   The best-fit values 
$\alpha_{0}$ and $\beta_{0}$  are given by minimizing a $\chi^{2}$ function,
changing $\alpha$ and $\beta$:
\begin{equation}
\chi^{2} = \sum_{i=1}^{6} \left(\frac{N^{\rm obs}_i - N^{\rm sim}_{i}(\alpha, \beta)}{\sigma^{\rm obs}_{i}} \right)^2,
\label{equ4}
\end{equation}
where $N^{\rm obs}_{i}$, $\sigma^{\rm obs}_{i}$ and 
$N^{\rm sim}_{i}(\alpha, \beta)$ are the observed number of excess counts,
its error and the number of remaining MC events after the analysis
assuming the spectrum $f(E) = \alpha E^{\beta}$, respectively, in the
{\it i}-th $\sum\rho_{\rm FT}$ bin among the six $\sum\rho_{\rm FT}$
bins between $10^{1.25}$ and 10$^{3.00}$ defined in $\S$\ref{s-3.1}.
In order to estimate $N^{\rm sim}_{i}(\alpha, \beta)$ in the same way
as experimental data, simulated secondary particles are inputted to
the detector response simulation.  Then, we obtain the expected
$N^{\rm sim}_{i}(\alpha, \beta)$ for the {\it i}-th $\sum\rho_{\rm FT}$ 
bin after the event reconstruction and event selections in the
same way as experimental data.  Here, the expected 
$N^{\rm sim}_{i}(\alpha, \beta)$ includes the energy resolution 
effect by the detector response simulation.

Subsequently, the differential $\gamma$-ray flux for the {\it i}-th 
$\sum\rho_{\rm  FT}$ bin is calculated by the following equation:
\begin{equation}
f_{i}(E_{i}) = \frac{N^{\rm obs}_{i}}{N^{\rm sim}_{i}(\alpha_{0}, \beta_{0})}~\frac{N_{\rm all}^{\rm sim}(\alpha_{0}, \beta_{0})}{\displaystyle 
\int_{E_{\rm min}^{\rm sim}}^{\infty} E^{\beta_{0}}~dE}~\frac{E_{i}^{\beta_{0}}}{~S_{\rm sim}~T_{\rm obs}},
\end{equation}
where $N_{\rm all}^{\rm sim}(\alpha_{0}, \beta_{0})$ denotes the total
number of MC events generated at the top of the atmosphere along one
diurnal motion assuming the spectrum $f(E) = \alpha_{0} E^{\beta_{0}}$, 
$\int_{E_{\rm min}^{\rm sim}}^{\infty} E^{\beta_{0}}~dE$ 
is the normalization factor of 
$N_{\rm all}^{\rm sim}(\alpha_{0}, \beta_{0})$, $E_{\rm min}^{\rm sim}$ 
denotes the minimum energy of simulated air shower events (0.3~TeV), 
$S_{\rm sim}$ denotes the area of core location distribution by the simulation
(300~m $\times$ 300~m $\times$ $\pi$), $T_{\rm obs}$ denotes the
observation live time, and $E_{i}$ denotes the representative energy
defined as the logarithmic mean of the energy calculated by the MC
simulation for the {\it i}-th $\sum\rho_{\rm FT}$ bin.

Figure~\ref{fig13} shows the differential energy spectrum of the
Crab~Nebula observed by the Tibet-III array together with the spectra
obtained by IACTs, including Whipple \cite{Hil98}, HEGRA \cite{Aha04},
CANGAROO~III \cite{Eno06},\\ H.E.S.S. \cite{Aha06} and MAGIC
\cite{Alb08}.  The differential flux for each $\sum \rho_{\rm FT}$ bin
is presented in Table~\ref{tbl-1}.  Finally, this energy spectrum is
fitted by the least $\chi^{2}$ method assuming $f(E) = \alpha
(E/3~{\rm TeV})^{\beta}$, and then we obtain the differential
power-law spectra as $(dJ/dE) = (2.09\pm0.32)\times10^{-12} (E/{\rm
  3~TeV})^{-2.96\pm0.14} {\rm cm}^{-2} {\rm s}^{-1} {\rm TeV}^{-1}$ in
the energy range of 1.7~TeV to 40~TeV.  
Note that the absolute
energy scale error in the Tibet-III array is experimentally estimated
to be smaller than $\pm$12\% by the moon's shadow observation
described in $\S$\ref{s-3.4}.  The energy scale uncertainty
corresponds to $(-28/+46)\%$ in the absolute $\gamma$-ray flux,
assuming the spectral index $-$2.96, which is our best-fit value.  
Our energy spectrum in this work
is consistent with other observations made by IACTs, such as HEGRA and
H.E.S.S., in the same energy range between 1.7~TeV and 40~TeV.

The previous flux measurement \cite{Ame99a}, with the Tibet-HD array
of 5,175~m$^2$ and an effective running time of 502.1 live days, is
approximately double this measurement.  In order to properly estimate
the difference between the previous work and the present one, we give
a re-fit to both data points from 2.8~TeV to 20 TeV in the overlapping
energy region assuming a power-law spectrum.  The previous (Tibet-HD)
and present (Tibet-III) energy spectra are expressed as
$(dJ/dE) = (5.04\pm0.94)\times10^{-12} (E/{\rm 3~TeV})^{-2.85\pm0.20} {\rm cm}^{-2} {\rm s}^{-1} {\rm TeV}^{-1}$ and
$(dJ/dE) = (2.35\pm0.49)\times10^{-12} (E/{\rm 3~TeV})^{-3.00\pm0.25} {\rm cm}^{-2} {\rm s}^{-1} {\rm TeV}^{-1}$,
respectively. The flux and spectral index differences between them are estimated to
be (2.69$\pm$1.06) $\times$10$^{-12} {\rm cm}^{-2} {\rm s}^{-1} {\rm
TeV}^{-1}$ and 0.15$\pm$0.32, respectively. As a result, the
combined statistical deviation between them is calculated to be
$\sqrt{(2.69/1.06)^2+(0.15/0.32)^2} \sigma = 2.6 \sigma$. Although we have
updated the MC simulation in this analysis, we cannot find any
systematics to explain this difference.  Hence, we conclude that the
higher flux observed in our previous measurement may have been caused
by a statistical signal fluctuation.

\subsection{Time Variability}

\begin{figure}[b]
\epsscale{1.20}
\plotone{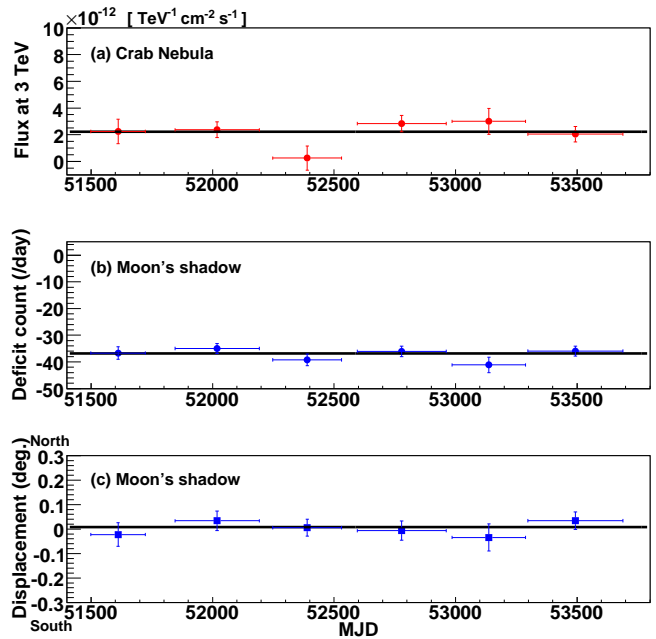}
\caption{Time variability of the Crab~Nebula and the moon's shadow
  observed by the Tibet-III array with $\sum\rho_{\rm FT} >10^{1.25}$
  between 1999 November and 2005 November. (a): Differential flux of
  the Crab~Nebula at 3~TeV.  (b): Daily deficit event rate averaged
  over one phase of the moon's shadow. (c): North-south displacement of
  the moon's shadow.}
\label{fig14}
\end{figure}





\begin{table}[b]
\begin{center}
\caption{Definition of six phases from 1999 November to 2005 November
  as shown in Fig.~\ref{fig14}. \label{tbl-2}}
\begin{tabular}{ccc}
\tableline\tableline
Phase & Period & Live Time \\
      &        & [days] \\
\tableline
1 & Nov. 18, 1999 - Jun. 29, 2000 & 173.1 \\
2 & Oct. 28, 2000 - Oct. 11, 2001 & 283.7 \\
3 & Dec. 05, 2001 - Sep. 15, 2002 & 201.8 \\
4 & Nov. 18, 2002 - Nov. 18, 2003 & 259.1 \\
5 & Dec. 14, 2003 - Oct. 10, 2004 & 123.6 \\
6 & Oct. 19, 2004 - Nov. 15, 2005 & 277.6 \\
\tableline
\end{tabular}
\end{center}
\end{table}



We divided our dataset from 1999 November to 2005 November into six
phases, as summarized in Table~\ref{tbl-2}, to examine the time
variability of the flux intensity.  Each phase corresponds to
approximately one calendar year.  We used slightly different
calibration parameters for each phase, because we usually calibrate
the scintillation detectors of the Tibet air shower array late in the fall of 
every year. Unfortunately, some of the blank periods seen in Table~\ref{tbl-2}
mostly coincide with the detector calibration periods, periods 
in which the air shower array was upgraded or when the data acquisition system 
was experiencing problems. The upper panel (a) in Figure~\ref{fig14} shows the
time variability of the $\gamma$-ray fluxes from the Crab~Nebula at
3~TeV. We found no evidence for the time variability of flux intensity
from the Crab~Nebula, as we can give a good $\chi^{2}$ fit to these
fluxes by a constant function ($\chi^{2} / d.o.f. = 6.55 / 5$), where
$d.o.f.$ means degrees of freedom.  In order to check the possible
systematics, the time variability of the deficit event rates of the
moon's shadow is also demonstrated as shown by the middle panel (b) in
Figure~\ref{fig14}.  The deficit event rates of the moon's shadow from
1999 November to 2005 November are very stable within a statistical
error $\pm$6\% year by year.  A fitting to the daily deficit event
rate averaged over a phase assuming a constant function is consistent
with a flat hypothesis ($\chi^{2} / d.o.f.  = 4.82 / 5$).  The lower
panel (c) in Figure~\ref{fig14} shows the time variability of the
north-south displacement of the moon's shadow, which is a reference to
the absolute pointing error described in $\S$\ref{s-3.2}.  It is also
very stable within $\pm0\fdg04$ during our observation period, and is
consistent with a flat hypothesis ($\chi^{2} / d.o.f. = 2.09 / 5$).
These systematics, estimated from observations of the moon's shadow, are
obviously negligible in the Crab~Nebula observation.

\subsection{Search for $\gamma$-Rays from the Crab Pulsar}

The rotation period of the Crab pulsar is 33~ms, as inferred from
radio, optical and X-ray observations.  A pulsed emission with that
rotation period at the GeV energy region has been detected by EGRET 
on board the CGRO satellite \cite{Fie98}, whereas several observations
have reported no evidence for pulsed emissions greater than 10~GeV
\cite{Les00,Nau02,Aha04,Aha07,Alb08}.  The emission models of
high-energy pulsed $\gamma$-rays are mostly based on the outer gap
\cite{Che86} and the polar cap \cite{Dau82} models.  The upscattered
pulsed $\gamma$-ray flux is also calculated by the inverse-Compton
process and the photon-photon absorption process assuming infrared
photon field models.  A model predicts an excessive flux around 1 $\sim$ 
10~TeV, depending on the infrared photon field models \cite{Hir01}.
Here, we present a search for pulsed $\gamma$-rays from the Crab
pulsar at energies from a few TeV to 100~TeV using the Tibet-III
array.

The arrival time of each event is recorded using a quartz clock
synchronized with GPS, which has a precision of 1~$\mu$s.  For the
timing analysis, all arrival times are converted to the solar system
barycenter frame using the JPL DE200 ephemeris \cite{Sta82}.  The Crab
pulsar ephemeris is calculated using the Jodrell Bank Crab Pulsar
Monthly Ephemeris \cite{JBC0,JBC1}.  The corrected arrival time of
each event is calculated to the rotation phase of the Crab pulsar, which
takes into account of the period derivative $\dot{P}$ of the period $P$ month by
month.

\begin{figure}
\epsscale{1.20}
\plotone{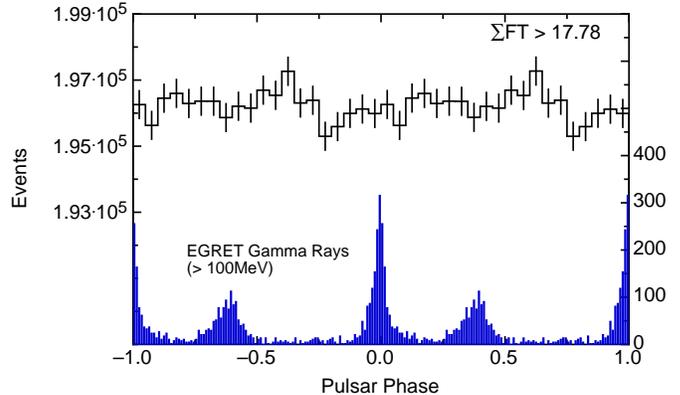}
\caption{ Distribution of the event phase of the Crab pulsar.  Phase 0
  is defined using the timing solution derived from the main pulse of
  the radio observations.  Upper plot shows our result for events with
  $\sum \rho_{\rm FT} > 10^{1.25}$ ($>$ $\sim$1~TeV).  Lower plot
  shows the $\gamma$-ray phase histogram above 100~MeV, as measured by
  EGRET \cite{Fie98}.}
\label{fig15}
\end{figure}

\begin{figure}
\epsscale{1.20}
\plotone{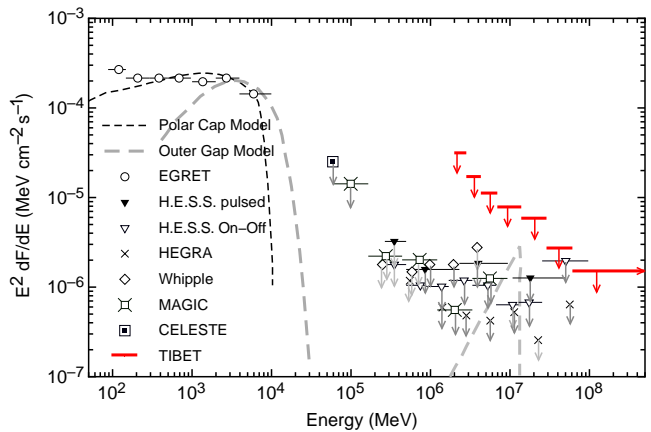}
\caption{Upper limits on the pulsed $\gamma$-ray flux from the Crab
  pulsar observed by the Tibet-III array (arrows with thick solid
  line), together with results from Whipple \cite{Les00}, CELESTE
  \cite{Nau02}, HEGRA \cite{Aha04}, H.E.S.S. \cite{Aha07} and MAGIC
  \cite{Alb08}. The long-dashed curve and dashed curve represent the
  fluxes expected from the outer gap and polar cap models,
  respectively.}
\label{fig16}
\end{figure}

Figure~\ref{fig15} shows the distribution of events for each phase in
two rotation periods of the Crab pulsar.  The distribution is
consistent with a flat distribution ($\chi^2/d.o.f. = 18.1/19$).
No significantly pulsed signal is found in observations for
events with $\sum \rho_{\rm FT} > 10^{1.25}$ ($>$ $\sim$1~TeV).  The
phase analysis is performed for each $\sum \rho_{\rm FT}$ bin to
examine the energy dependence.  Table~\ref{tbl-3} shows the
statistical results by the $Z^2_2$ test \cite{Buc83}, $H$ test
\cite{Jag94} and least $\chi^2$ test.  Almost all the statistical
tests show that the phase distributions are uniform within a 3~$\sigma$
significance level.  We estimate the 3~$\sigma$ flux upper limit
on the pulsed emission from the Crab pulsar using the $H$ test
\cite{Jag94} as
\begin{eqnarray}
\nonumber
x_{3\sigma} & = & (1.5+10.7\delta)(0.174H)^{0.17+0.14\delta}\\
\nonumber
 & & \times  \exp \{(0.08+0.15\delta) \\
 & & \times (\log_{10}(0.174H))^2\},
\end{eqnarray}
where $\delta$ is the duty cycle of the pulsed component, assuming
$\delta$ is 21\% for the Crab pulsar.  Exposure from the Crab pulsar
to the Tibet-III array is estimated using MC simulation, assuming
the differential energy spectrum for $\gamma$-ray emission has a
spectral index of $-$2.6.  Upper limits are compared to previous
results inferred from other experiments, as shown in
Figure~\ref{fig16}.

\begin{table}
\begin{center}
\caption{Results of statistical tests for pulsed emissions.  $\chi^2$-,
  $Z^2_2$- and $H$-test (probabilities) are calculated for a flat
  phase distribution.
\label{tbl-3}}
\begin{tabular}{cccc}
\tableline\tableline
$\sum \rho_{\rm FT} $ & $\chi^2/d.o.f.$ & $Z^2_2$ & $H$ \\
\tableline
$10^{1.25}$ -- $10^{1.50}$ & 0.97 (0.49)   & 9.62 (0.047) &  9.62 (0.021)\\
$10^{1.50}$ -- $10^{1.75}$ & 1.21 (0.24) & 7.64 (0.11) &  7.64 (0.047)\\
$10^{1.75}$ -- $10^{2.00}$ & 0.81(0.70)  & 2.54 (0.64) & 4.49  (0.17) \\
$10^{2.00}$ -- $10^{2.33}$ & 0.35 (0.96) & 2.30 (0.68) & 6.14  (0.086) \\
$10^{2.33}$ -- $10^{2.67}$ & 1.41 (0.11) &  9.68 (0.046) &  14.56 (0.0030) \\
$10^{2.67}$ -- $10^{3.00}$ & 0.80 (0.71) & 3.67 (0.45) & 6.09  (0.088) \\
$>$ $10^{3.00}$  & 0.60 (0.91) & 1.11 (0.89) & 4.48  (0.17) \\
\tableline
$>$ $10^{1.25}$  & 0.95 (0.52) & 8.41 (0.078) & 8.87  (0.029) \\
\tableline
\end{tabular}
\end{center}
\end{table}



\section{SUMMARY AND PROSPECTS}

We have been successfully operating the Tibet-III air shower array at
Yangbajing in Tibet, China since 1999. Using the dataset collected by this array from 1999
November through 2005 November, we obtained the
differential energy spectrum of $\gamma$-rays from the Crab~Nebula as
$(dJ/dE) = (2.09\pm0.32)\times10^{-12} (E/{\rm 3~TeV})^{-2.96\pm0.14}
{\rm cm}^{-2} {\rm s}^{-1} {\rm TeV}^{-1}$ in the energy range of
1.7~TeV to 40~TeV. This result is consistent with data
obtained by IACTs, and is statistically consistent with our
previous result within 2.6~$\sigma$.  No evidence is found for time
variability of flux intensity from the Crab~Nebula at multi-TeV
energies in comparison with the long-term stability of the moon's
shadow.  We also searched, unsuccessfully, for pulsed $\gamma$-rays from the Crab
pulsar at multi-TeV energies.

In this paper, we have carefully analyzed the moon's shadow observed with
the Tibet-III array to calibrate the energy of primary cosmic rays
directly.  In general, this energy is indirectly estimated by
measuring shower size in air shower experiments. 
The cosmic-ray beams coming
from the moon's direction are bent by the geomagnetic field, so
that the moon's shadow should shift to the west depending on the
primary energy.  We tried to directly estimate the primary energy by measuring
the displacement of the moon's shadow.  
This energy scale estimation is the first attempt and obtained that the
systematic error in the absolute energy scale observed by the
Tibet-III array is estimated to be less than $\pm$12\% at energies
around 10~TeV.  The array's systematic pointing error is also
estimated to be smaller than $0\fdg011$. The long-term stability of
the deficit rate of the moon's shadow was within
a statistical error $\pm$6\% year by year, thus confirming the
stability of the array operation. This calibration method is very
unique and will be important to ground-based TeV $\gamma$-ray
observations.

In the near future, we will set up a 10,000~m$^{2}$
water-Cherenkov-type muon detector (MD) array in the ground beneath the
Tibet air shower (AS) array \cite{Ame07b,Ame07c,Ame07d}.  
This Tibet MD array will significantly improve
$\gamma$-ray sensitivity of the Tibet air shower array above 10~TeV by
means of $\gamma$/hadron separation based on counting the number of
muons accompanying each air shower.  The energy spectrum of the
Crab~Nebula can be surely measured up to several hundred TeV, if extended,
with a low background level, using the Tibet AS+MD array. This new array
will enable us to survey not only the known sources but also new
sources in the northern sky above 10~TeV, and may be superior to
IACTs for observing diffuse $\gamma$-ray sources \cite{Ame06,Abd07} and
diffuse $\gamma$-rays from the galactic plane \cite{Ame02,Atk05},
owing to its wide field of view and high rejection power for
hadronic showers.


\acknowledgments

The collaborative experiment of the Tibet Air Shower Arrays has been
performed under the auspices of the Ministry of Science and Technology
of China and the Ministry of Foreign Affairs of Japan. This work was
supported in part by a Grant-in-Aid for Scientific Research on Priority
Areas from the Ministry of Education, Culture, Sports, Science and
Technology, by Grants-in-Aid for Science Research from the Japan Society 
for the Promotion of Science in Japan, and by the Grants 
from the National Natural Science Foundation of China and the
Chinese Academy of Sciences.

\end{document}